# AZ Model for Software Development

Ahmed Mateen
Department of
Computer Science,
University of Agriculture
Faisalabad,
Pakistan

Muhammad Azeem Akbar
Department of
Computer Science,
University of Agriculture
Faisalabad,
Pakistan

Mohammad Shafiq
Department of
Computer Science,
University of Agriculture
Faisalabad,
Pakistan

## ABSTRACT
Know a day's Computer system become essential and it is most commonly used in every field of life. The computer saves time and use to solve complex and extensive problem quickly in an efficient way. For this purpose, software programs are develop to facilitate the works for administrator, offices, banks etc. so Quality is the most important factor as it mostly defines CUSTOMER SATISFACTION which directly related to success of the project, so there are many approaches (methodologies) have been developed for this purpose occasionally. The main study of this paper is to propose a new methodology for the development of the software which focuses on the quality improvement of all kind of product. This study will also discuss the features and limitation of the traditional methodologies like water-fall, iterative, spiral, RUP and Agile and show how the new innovative methodology is better than previous one.

## Keywords
Software process model, high quality product, innovative methodology, Traditional Development Models, propose Model.

## 1. INTRODUCTION
Now a days, Software Engineering has become a growing and emerging field in the world. Every field is relaying on software; software has made life comfortable. Nobody can deny the importance of the software. The quality of the software is key feature for success of the any system and the quality product is developed by using different kind of methodologies like Waterfall, spiral, RAD, RUP, Agile and etc. whereas the selection of the methodology depends on what type of the scenario in which product is being developed. The main purpose of these software development models is to develop software that fulfill the stakeholder's requirements purely within the prescribed budget and time. A successful project is that which satisfy the stakeholder requirements and should be completed within the time and budget.

### 1.1 Waterfall model
In software development models water fall is one of the oldest and commonly known model to develop the systems.it is commonly used for developing the projects of both government and public limited companies. The importance of waterfall model is that it is a sequential model. All the phases of this model should be followed sequentially and in downward. The phases are requirement gathering, analyses, design, coding, testing and maintenance.it is ensure that the design phase always be followed before the coding.in such model the design flows before the development of the product. This model is commonly used for those product whose major concern is quality control. Due to its focused documentation and planning Requirement [1] engraining phases that construct this model in such a way that could not over lapped so the water fall model starts and end before the new product is being started. The described steps stretch a brief explanation about the waterfall process. Requirement gathering is the first and very complex phase because requirements of the systems are collected from the innocent Stakeholder. Hence, the agreement between the clients and the developers for the software specification is established. Hence the requirements are collected and analyzed and the proper documentation is completed this documentation is enable to help in the system development life cycle which helps further in the development process. Designing is the second phase of water fall model this phase depend upon the first phase which is requirements gathering. [1, 2]. When the requirement gathering phase is completed gathered information are evaluated for proper implementation. For the process solution it is the procedure of planning and problems solution. Appropriate algorithm, design and architecture should by chose by dealing that devilment model. Third is coding when whole the requirements are collected that will be coded in any programming language. Testing is the fourth phase of waterfall model in this phase requirement should be tested with the developed system and the existing errors are detected and fixed. Fifth phase of the waterfall model is Maintenance After releasing the software some little bit correction or editing is made according to the stakeholder's satisfaction.

### 1.2 Spiral Model
is a system development methodology that are combined with element of design and prototyping due to the combination of these stages advantages of top down and the bottom up should be taken. It's a meta-model which means to say it can be used with other developing models [1, 3]. The main focuses of this model is to assess and minimize risk. It can be attained by dividing a system into smaller blocks or smaller parts, that's provide the facility for ease modification during the development process also casual to detect risks and high deliberation of system continuance during the life cycle. The development team starts development with a small set of requirements and then drives from every development phase according to the requirements. It's a cause to learn about the expected bugs from the initial iteration (via a risk analysis process).and the development team will be able to made addition as a Non-functional requirement. When the system is completed and ready for the implementation and maintenance phase. Prototype playing the key effect for furnishing the application. For the description of Spiral model phases are described as follows: Always project is started with planning in this phase the requirements of the customer are understood for this understanding continuous communications are conducted between the system analysts and the customer.

After the planning Risk should be analyzed. This process is made to identify the risk and substitute solutions. Prototype is accompanied at the end of this phase. Development/ Engineering is being underway after risk analyses this phase include the software production and testing. After completing the development and testing evaluation is made. The result of the system is evaluated by the stakeholder before the system endures to the next round or next spiral [4, 5, 6, 7].

### 1.3 Incremental and iterative Model
Waterfall models phases are combined as an iterative manner. Moreover, a deliverable increments of the software are produces by each linear sequence.in the first increment phase basic requirements are gathered these requirements are the base of the product, therefore many additional features (known, unknown) remain undeliverable in this iteration. Partial implementation of the total system is constructed by this model. Then, gradually it adds increased functionality. However, each subsequent should be released by adding a function of the previous subsequent until the implementation of designed and functionalities [4, 5, 6, 7].

### 1.4 Rapid application development (RAD)
in programming system that provides the facility to the programmer to build a working program quickly. However, the numbers of tools are providing by the RAD system that help to build graphical use interface normally taken by a large efforts of development. RAD system is the most known for windows are Visual basic and Delphi. By using the RAD system, the devolving time and budget should be reduced and the efficient executable code should be generated. Nowadays, by using the RAD systems extremely fast code should be produced. Traditionally many traditional environments for programming a number of visual tools to aid development. However, the difference between other development environments system and rad RAD has become blurred [8].

### 1.5 Rational unified process model
Rapid development is used in the field of Web applications from past ten years. For the better development of Web application, there is a major need to somehow model it, before this can actually develop it in real life. For the Modeling of Web applications, it is not only adding some description as text or drawing nice pictures. Itself this process needs certain methodology and very complex, not only using the traditional software development processes, but reaction time to changes and also creative mind and short. The Web Applications has reliable architecture and visually attractive and desirable user interfaces as well. Usually the services are provided by using the three layers a data layer, an application layer and a client. It is a common known that the users are irritated. However, if the Web applications design is poor and the information is not up to date, commonly it's the main cause to lose the customer because of frustration. A powerful media covers by Web applications [9, 10]

## 2. METHODOLOGY AND TECHNIQUE OF PROPOSED MODEL
In introduction phase it has seen the weakness of the waterfall, agile, spiral, RUP and RAD development life cycle these all model are best to develop product in different scenario its domain is limited. In this study author has proposed a model which is the best for all scenarios and give the maximum outputs to produce high quality products. The time constraints are a very important factor in software development life cycle. The proposed model contains the time boxing parameter that enable the project manager to complete the project in time and deliver the system to the customer according to the contract. The intima delivery is an goodwill affection of the software house and good reputed of project manager the proposed model is shown blow in Which time boxing parameter are applied with important phases.

I. **Communication:** Communication is the bridge between the stakeholders and the requirement engineers to elicitate the true requirement. So the prototyping methodology will also be used to elicitate the requirement from the stakeholders.

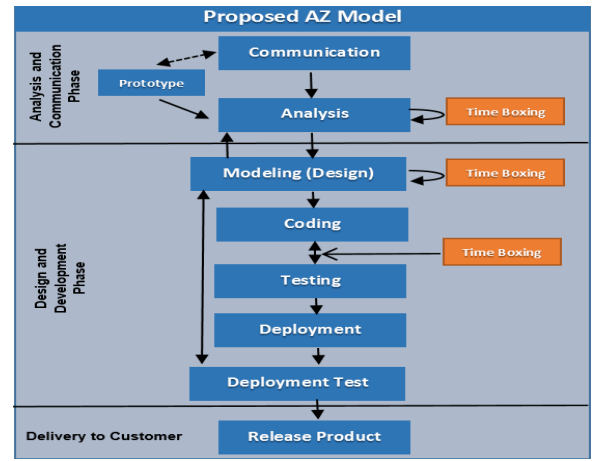

**Fig.1: Innovative model for Quality enhancement**

II. **Prototype:** It is visual representation of the requirements. It may be any format (textual, animated, audio, video, graphical or some model) which provide the user friendly environment to elicitate the true requirement of the system.

III. **Analysis:** It focuses on describing function, behaviors and required data which can be viewed for completeness, correctness, consistency. Moreover, it reflects the need of all stakeholders

IV. **Design:** Design provide the virtually description to every engineer about detail model of the software interfaces, architecture, and the components that are necessary to implement the system

V. **Coding:** In computer, **source code** is the instruction written in high level language (possibly with comments) written using some human-readable computer language

VI. **Testing:** Testing is a process in which the results of the systems or units are recorded and an evaluation is made of some aspect of the system or component

VII. **Deployment:** In this model deployment phase deals with delivery, support and feedback.

VIII. **Release product**: it is the last phase of the AZ model in which finely product handed over the customer.

### 2.1 Features of proposed Model
The working of the model is divided in to four phases prototyping phase is ensure to get clear requirements which are essential to produce high quality product and time boxing is used for in time delivery of the software and rest of the two phases are purely deal with the quality enhancement. Which are following.

- Parallel testing and code
- Feedback and iteration
- Prototyping

- Tim boxing

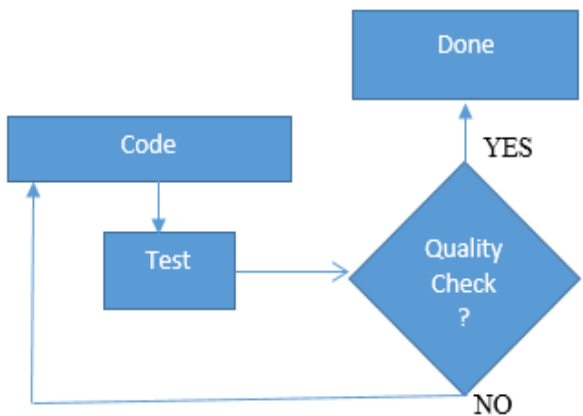

**Fig. 2: Parallel testing**

To produce a successful system, the requirement must be very clear and understandable so that it fulfills all requirements of the customer that's accepted from the system. There are so many techniques to elicitate the requirement but prototyping in one of the best technique to elicitate the clear requirement of both (fictional and Non-functional). Prototyping is visual representation of the requirement. It may be any format (textual, animated, audio, video, graphical or some model) which provide the user friendly environment to elicitate clear requirement of the system. But it is clear that it can avoid the prototyping in rear cases it has been seen where the customer is very knowledge about the system and can state clearly what 'she wants. In this situation the prototyping is useless and time consuming. So this phase is best for all type of the scenario moreover very helpful to elicitate the clear requirement that is essential to produce high quality product.

## 2.2 Parallel testing and code
Testing is very important activity that finds errors and check the functionality of the system is satisfied or not. Recovering and fixing errors is more costly after releasing the product some time it causes to failure of the product. That's why in the model testing phase is introduced parallel with coding. Coding and testing is started at the same time. Code is done and test until the required quality not gain. Through this process the quality of the system enhanced.

## 2.3 Feedback and iteration
Feedback is also a very important phenomenon to improve the quality. It clears that what was the expectation of the user from the system. On the other hand, doing thing again and again is also helpful for improving the quality.

Feedback from one phase to another phase is to going back to previous phase. From fig (1) it can be viewed that this can return to previous phase as from modeling to analysis phase. Customers are always motivated during throughout the project and involved where needed.

Iteration means remaining the same phase that is doing repeated cycle analysis, design and testing parallel with the coding. It is more easy and cost effective doing repeated cycle remaining on the same phase means reanalysis, redesigning and retesting than feedback to previous phase. Doing the iterations, the management play an important rule to keeping track all the schedule, the customer feedback is the base of the iterations or finding error or to improve quality of (structure, efficiency, usability, reliability, efficiency, and achieving goals) product.

## 2.4 Prototyping
Requirements are the essential and first phase of system development life cycle. Prototyping play an important role when the stakeholder unaware from the product or It can say if the customer or stakeholder not able to convey the pure requirements. the prototyping will be used which visualized the product and its related feature by seeing the prototype product customer high light the changes according to the requirements.so the prototyping phase made enable to collect the true requirement.

## 2.5 Time boxing
As the other approaches of iterative development use the time boxing. Therefore, as in iterative developed the working system is delivered after completing every iteration. Therefore, in time boxing the total length of the time is divided into the appropriate time duration which is require for completing each iteration. The units of time which is provided to an iteration is a time box, which is static duration for completing the said iteration in the said time box all activities of the said iteration that's needed to be performed for successfully release the developed iteration. So, the period is Static, the main factor of selection the time box is that to build the units or system feature and requirements within the selected time box. Time boxing is made according to nature of the phase of the development model. Each phases performs some purely defined task of the iteration and produce the output according to the requirement. The output of the one phase become the input of the next phase, therefore if the output of the current phase is correct then it made the correct input to the next phase. When the result is passing to the nest phase activity of development passed over to the next phase. Note that giving the over will need of some objects to be created by a phase which provides all the information that is needed to perform the next phase. However, by using the time boxing technique it an able to the project manager to define the appropriate time for each phase in which all the activities should be completed in figure (1) the time box is shown that define the fixed time for each phase.

## 3. RESULTAS OF PROPONED MODEL
The innovative methodology for development of software would be the best for all kind of scenario to produce high quality product within the time duration. In the following propose model is compare with the traditional model. SDLC is a methodology which includes Requirement engraining, designing, building, and maintaining information and industrial systems. However, there is lot of SDLC models exist that are used for devolving the software's, such as Waterfall model, which is distributed into different parts to be completed consecutively for devolving a project, such as Spiral model is also used for developing the systems, in which a project is fleeting through some number of iterations. Another model which is called incremental model or incremental desisgn.it contain seven phases which is planning, requirement engraining analysis design implementation deployment testing and evaluation. Therefore, many researchers investigate the weakness and strength of the SDLC. Such as Waterfall, spiral, incremental, rational unified process (RUP), rapid application development (RAD), agile software development, and rapid prototyping are some successful models. However, all suggested SDLC model describe the basic properties. They all ordered the phases or stages in a sequence. For achieving and completing the success full project that prescribed sequence should be followed. However, in this Study strengths and weaknesses of all the defined developing models are discussed. And the comparison is made as follows.

**Table 1. Proposed AZ model objective**

| Traditional Model | Drawback | Solution by Propose AZ Model |
|---|---|---|
| Water Fall | Requirement Document Freeze. There is no change once document final and restricted the customer involvement | Requirement document does not freeze can be change where necessary. This model includes the concept of prototyping which give the maximum user involvement. |
| Prototyping | This model essential condition base on to get requirement through prototyping. | In proposed model there is a direct way to go analysis phase without prototyping |
| RAD | Rapid application development produces low quality product as its not consume so much time on requirement gathering, Analysis and designing rather than its focuses on Development phases | In the model much more focuses on the Requirement, analysis and designing as well as coding testing and deployment. |
| RUP | Rational unified process for development only focuses on functional requirement | This propose model deal with functional and non-functional requirement |
| Spiral | Spiral methodology has not time limit to get a mile stone and require very experience project manager and team | This methodology depends on dead line according to every process. Less experience team and project manager are sufficient for this |
| Agile | Agile process purely depends on the team work and your customer should be very experience, knowledgeable | Propose model developed for all kind of customer. This model does not depend on the team work every one can do their jobs separately |

## 4. CONCLUSION AND FURTURE WORK

The main idea of the paper is to introduce a new innovative methodology that overcomes weakness in traditional methodology (Waterfall, RUP, Agile, Spiral and RAD). In this new model, the testing starts with coding and proceeds in parallel. To get clear and true requirement the prototype phase is introduced in the model, moreover Feedback between phases and iterations on the same phase is the main idea to produce high quality product. This is the main purpose to introduce the new methodology. In future it is suggested to apply this model in different projects according to its mechanical processes as well as simulate the architectural design for its proper working and explanation with documentation so the diversity of model will be described which will be beneficiary to project managers for decision making throughout the development phases.